\def\rfr#1{eq. (\ref{#1})}

\def\pms{P^{(\rm meas)}}

\def\bar{\begin{eqnarray}}
\def\ear{\end{eqnarray}}
\def\bb{\bibitem}
\def\eqi{\begin{equation}}
\def\eqf{\end{equation}}
\def\eqia{\begin{eqnarray}}
\def\eqfa{\end{eqnarray}}
\def\rp#1#2{{#1\over#2}}

\def\lb#1{\label{#1}}

\def\oc2{$\mathcal{O}(c^{-2})$}

\documentclass[11pt]{article}
\usepackage{amsmath,amsthm,amscd,amssymb}
\usepackage{latexsym}
\usepackage{graphicx,epsfig}
\usepackage[edtable]{lineno}

\begin{document}
%\pagewiselinenumbers % ad ogni pagina riprende da 1 il conteggio delle linee
%\runninglinenumbers % conta le linee da 1 a X fino alla fine del file

\noindent{\bf \LARGE{Dynamical constraints on the quadrupole mass
moment of the HD 209458 star}}
\\
\\
\\
{Lorenzo Iorio}\\
{\it Viale Unit$\grave{a}$ di Italia 68, 70125\\Bari, Italy
\\tel./fax 0039 080 5443144
\\e-mail: lorenzo.iorio@libero.it}

\begin{abstract}
The aim of this paper is to dynamically constrain the quadrupole
mass moment $J_2$ of the main-sequence HD 209458 star. The adopted
method is the confrontation between the measured orbital period of
its transiting planet Osiris and a model of it. Osiris is assumed
to move along an equatorial and circular orbit. Our estimate, for
given values of the stellar mass $M$ and radius $R$ and by
assuming the validity of general relativity, is $J_2=(3.5\pm
385.1)\times 10^{-5}$. Previous fiducial evaluations based on
indirect, spectroscopic measurements yielded $J_2\sim 10^{-6}$:
such a value is compatible with our result.
\end{abstract}


\section{Introduction}
The quadrupole mass moment $J_2$ is an important astrophysical
stellar parameter related to the inner structure and dynamics of a
star (Patern\`{o} et al. 1996; Pijpers 1998).

In this paper we dynamically constrain the quadrupole of the
main-sequence star HD 209458 from the measured orbital period $P$
of its transiting planet HD 209458b `Osiris' (Charbonneau et al.,
2000; Henry et al., 2000). It is the best-studied transiting
planet to date, mainly due to its proximity and the resultant high
apparent brightness of its parent star.

Until now, only a fiducial value $J_2=2\times 10^{-6}$ by
Miralda-Escud$\acute{\rm e}$ (2002), based on spectroscopic
measurements of rotational velocity (Queloz et al. 2000), exists
for HD 209458. Also Winn et al. (2005) assumed $J_2\sim 10^{-6}$.
More generally, Miralda--Escud$\acute{\rm e}$ (2002) investigated
the possibility of dynamically measuring the quadrupole of a star
from the node and periastron precessions of a transiting planet
which cause a time variation of the duration of a transit; the
periastron precession also induces a variation of the transit
period. Such orbital perturbations should be measured from an
accurate photometric analysis of the light-curve of the transiting
planet, but, until now, such a proposed strategy has not yet been
implemented. Winn et al. (2005) pointed out that, for $J_2\sim
10^{-6}$, the quadrupole node precession would amount to about 4
arcseconds per year; measuring such an effect would require
high-precision photometry spanning several years (Winn et al.
2005).
\section{The use of the orbital period of Osiris}
The Newtonian gravitational potential $U$ of an oblate star of
mass $M$ and equatorial radius $R$ can be written as \eqi
U=-\rp{GM}{r}+\rp{GMR^2J_2}{2r^3}(3\cos^2\theta-1),\lb{upot}\eqf
where $\theta$ is the co-latitude angle.  For the orbital period
of a planet of mass $m$ in a circular and equatorial
($\theta=\pi/2$) orbit of radius $a$ \rfr{upot} yields
\eqi P^{(\rm N)}\equiv P^{(\rm
0)}+P^{(J_2)}=2\pi\sqrt{\rp{a^3}{G(M+m)}}-\rp{3\pi R^2
J_2}{2\sqrt{G(M+m)a}}.\lb{per} \eqf
In fact, in addition to \rfr{per}, there is also a post-Newtonian,
general relativistic part (Soffel 1989; Mashhoon et al. 2001;
Iorio 2005; 2006) to be added; for circular orbits and $m\ll M$
(see Section \ref{appe}) it is \eqi P^{(\rm
PN)}=\rp{3\pi\sqrt{G(M+m)a}}{c^2}\lb{gr},\eqf where $c$ is the
speed of light in vacuum, so that
\eqi P=P^{(0)}+P^{(J_2)}+P^{(\rm PN)}.\lb{uff}\eqf In the case of
Osiris, \rfr{per}-\rfr{gr} yield a reliable model of its orbital
period because the eccentricity $e$ was recently evaluated to be
$e=0.014\pm 0.009$ (Laughlin et al. 2005) and the inclination
angle $\psi$ of the orbital plane to the star's equator,
determined by means of the Rossiter-McLaughlin effect (Rossiter
1924; McLaughlin 1924), should not be larger than about 5 deg,
(Winn et al. 2005). Moreover, there are currently no observational
evidences of the presence of other bodies  around HD 209458 (Brown
et al. 2001; Croll et al. 2005; Laughlin et al. 2005; Agol \&
Steffen 2006) which may require the introduction of additional
perturbing terms in \rfr{uff}; the inclusion of the general
relativistic correction of \rfr{gr} is required because it amounts
to about 0.1 s, while the errors in the most recent measurements
of the Osiris' orbital period are 0.016 s (Wittenmyer et al. 2005)
and 0.033 s (Knutson et al. 2006).

Thus, the HD 209458 quadrupole mass moment can be determined by
comparing the model of \rfr{per}-\rfr{uff} to the measured period
$P^{(\rm meas)}$, determined in a purely phenomenologically way
from combined photometric transit and spectroscopic radial
velocity techniques, independent of any gravitation theory, and
solving for $J_2$
\eqi J_2=-\rp{2P^{(\rm meas)}}{3\pi R^2}\sqrt{G(M+m)a} +
\rp{4}{3}\left(\rp{a}{R}\right)^2+ \rp{2G(M+m)a}{c^2
R^2}\lb{oiu}.\eqf By using \rfr{oiu} and the system parameters
derived for $M=1.07$M$_{\odot}$ and $R=1.137$R$_{\odot}$ and
$P^{(\rm meas)}=3.52474554$ d (Wittenmyer et al. 2005), we obtain
\eqi J_2=3.5\times 10^{-5}.\lb{j2}\eqf

It must be noted that the obtained result is free from any a
priori, `imprinting' effect by $J_2$ itself. Indeed,  $M$ and $R$
are kept fixed,
%the impact of
% $m$, which is, in fact, obtained from the mass function
% with a simple Keplerian model of the period $P^{(0)}$, is negligible (see \rfr{j2err}
% below),
 and $a$, determined from\footnote{$K_M=\left(\rp{2\pi a_M}{P}\right)\sin i=\left(\rp{m}{m+M}\right)\left(\rp{2\pi a}{P}
 \right)\sin i$ is the projected semiamplitude of the star's radial velocity.}
 \eqi \rp{K_M^3 P^3}{8\pi^3}=\rp{a^3m^3\sin^3 i}{(m+M)^3}\eqf
 which is independent of any model of the orbital period, is not affected by $J_2$ over timescales longer than
one full orbital revolution.

 Let us now evaluate the uncertainty in $J_2$ as
\eqi\delta J_2\leq \delta J_2^{(P) }+ \delta J_2^{(a)} + \delta
J_2^{(m)}.\eqf For the same values of $M$ and $R$ as before and
$\delta\pms=0.016$ s (Wittenmyer et al. 2005) we have
\begin{equation}\left\{\begin{array}{lll}
\delta J_2^{(a)}  =  \left[ \rp{8}{3}\rp{a}{R^2} + \rp{P^{(\rm
meas )}}{3\pi R^2} \sqrt{ \rp{G(M+m)}{a} } + \rp{2G(M+m)}{c^2 R^2
}\right]\delta a=2.404\times
10^{-3},\\\\
\delta J_2^{(m)}  =  \left[ \rp{P^{(\rm meas )}}{3\pi
R^2}\sqrt{\rp{Ga}{M+m}} + \rp{2Ga}{c^2 R^2}\right]\delta
m=1.437\times 10^{-3},\\\\
\delta J_2^{ (P) }=\left[\rp{2\sqrt{G(M+m)a}}{3\pi
R^2}\right]\delta P^{(\rm meas)}=5\times
10^{-6}.\lb{j2err}
\end{array}\right.\end{equation}
%\delta J_2^{(G)}  =  \left[ \rp{P^{(\rm meas )}}{3\pi
%R^2}\sqrt{\rp{(M+m)a}{G}} + \rp{2(M+m)a}{c^2 R^2}\right]\delta
%G=\times 10^{-}.\lb{j2err}\end{array}\right.\end{equation}
%
%
The stellar mass was not included in  the least-square solution by
Wittenmyer et al. (2005) also because its determination is more
model-dependent than the other parameters. The range of allowable
values $1.06 \pm 0.13$ solar masses (Cody \& Sasselov 2002) was,
instead, used; it comes from observational errors in temperature,
luminosity, and metallicity as well as systematic errors in
convection mixing-length and helium abundance. The resulting
scattering in the determined values of $J_2$  is \eqi 3.2\times
10^{-5}<J_2<3.7\times 10^{-5}.\eqf Thus, we can state that the
model-dependence of our estimate amounts to $5\times 10^{-6}$, so
that a conservative estimate of the total uncertainty in $J_2$ is
\eqi \delta J_2\leq 3.851\times 10^{-3}.\eqf

\section{Discussion and conclusions}
In this paper we dynamically constrained the quadrupole mass
moment $J_2$ of the HD 209458 star from the orbital period $P$ of
its transiting planet Osiris, assumed to be in a circular and
equatorial orbit. Its measured value$-$determined in a
phenomenological way, independent of any gravitational
theory$-$was compared to an analytical model including the
Newtonian part, constituted by the usual Keplerian component and
the term induced by $J_2$, and the post-Newtonian, general
relativistic correction. The inclusion of the latter term, which
is of the order of 0.1 s, is motivated by the $\sim 0.01$ s level
of accuracy reached in measuring $P$. By keeping the stellar mass
$M$ and radius $R$ fixed to values within a range determined from
stellar evolution models and temperature/luminosity measurements,
by assuming that general relativity is valid in the HD 209458
system as well and that Osiris is the only planet affecting the
motion of its parent star in a detectable way, we obtain
$J_2=(3.5\pm 385.1)\times 10^{-5}$. While the uncertainty due to
the error in the orbital period amounts to $\sim 10^{-6}$ only,
the Osiris' mass and semimajor axis boost the total bias to $\sim
10^{-3}$. Previous fiducial evaluations based on indirect,
spectroscopic measurements giving $J_2\sim 10^{-6}$ are compatible
with our result.

In order to make easier a comparison with our results, in Table
\ref{numval} we quote the numerical values used for the relevant
constants entering the calculation.

\begin{table}
\caption{ Values used for the defining, primary and derived
constants
(http://ssd.jpl.nasa.gov/?constants$\#$ref).}\label{numval}

\begin{tabular}{llllll} \noalign{\hrule height 1.5pt}
constant & numerical value & units & reference\\
$c$ & 299792458 & m s$^{-1}$ & {\small(Mohr \& Taylor 2005)}\\
$G$M$_{\odot}$  & $1.32712440018\times 10^{20}$ & m$^3$ s$^{-2}$ &  {\small(Standish 1995)}\\
$G$ & $(6.6742\pm 0.0010)\times 10^{-11}$ & kg$^{-1}$ m$^3$ s$^{-2}$ &  {\small(Mohr \& Taylor 2005)}\\
R$_{\odot}$ & $6.95508\times 10^8$ & m &  {\small(Brown \& C.-Dalsgaard 1998)}\\
%AU & $1.49597870691\times 10^{11}$ & m & (Standish 1998)\\
$1$ mean sidereal day & $86164.09054$ & s &  {\small(Standish 1995)}\\
\hline

\noalign{\hrule height 1.5pt}
\end{tabular}

\end{table}
%

%An independent check of our
%result may come from measurement of the node and apsidal line
%precessions, although they should require high-precision
%photometry spanning many years.
\begin{appendix}
\section{The post-Newtonian correction to the orbital
period}\lb{appe}
In fact, the post-Newtonian gravito-electric correction to the
orbital period does depend on both the eccentricity $e$ and the
initial value of the true anomaly $f_0$ according to \eqi P^{(\rm
PN )}=\left[\rp{3\pi}{c^2}\sqrt{G(M+m)a}\right]F(e,f_0),\eqf with
(Soffel 1989; Mashhoon et al. 2001) \eqi
F(e,f_0)=3-\rp{\nu}{3}-\rp{2\sqrt{1-e^2}}{(1+e\cos f_0)^2}, \eqf
and \eqi\nu=\rp{mM}{(M+m)^2}.\eqf Depending on $e$ and $\nu$,
$F$--and $P^{(\rm PN)}$--vanishes for those values of $f_0$ which
satisfy the relation \eqi \cos
f_0=\rp{1}{e}\left[\sqrt{\rp{6\sqrt{1-e^2}}{9-\nu}}-1\right];\lb{form}\eqf
 it may also happen that the absolute value of the
right-hand-side of \rfr{form} is larger than 1, so that $P^{(\rm
PN)}\neq 0$.

In the case of the HD 209458 system, $\nu\sim 10^{-4}$; recent
refinements in the Osiris ephemerides yields $e\leq 0.023$
(Laughlin et al. 2005), so that \eqi
\rp{1}{e}\left[\sqrt{\rp{6\sqrt{1-e^2}}{9-\nu}}-1\right]\sim
-8:\eqf $F$ never vanishes, ranging from 0.90 to 1.09. Thus, the
difference between the maximum and the minimum values of $P^{(\rm
PN)}$ is 0.019 s at the most: it just lies at the edge of the
precision with which the orbital period is known, i.e. $ 0.016$ s
(Wittenmyer et al. 2005) and $0.033$ s (Knutson et al. 2006), so
that we can approximate $F$ to unity.
\end{appendix}

%-----------------------------------------

%-------------------------------------

\end{document}